\documentclass[12pt,thmsa]{article}
\usepackage{amsfonts}
\usepackage{amsmath}
\usepackage{amssymb}

\input{tcilatex}
\begin{document}

\title{Abstract Harmonic Analysis on Spacetime. }
\author{Kahar El-Hussein \\
Department of Mathematics, Faculty of Science, \\
Al Furat University, Dear El Zore, Syria \ \ and\\
Department of Mathematics, Faculty of Arts Science Al Quryyat, \\
Al-Jouf University, KSA \\
\textit{E-mail kumath@hotmail.com }}
\maketitle

\begin{abstract}
Let $G=SL(2,%
\mathbb{C}
)$\ be the $2\times 2$ connected complex Lie group and let $P=\mathbb{R}%
^{4}\rtimes SL(2,$ $%
\mathbb{C}
)$ be the Poincare group (space time). In mathematics, the Poincar\'{e}
group (spacetime), named after Henri Poincar\'{e}, is the group of
isometries of Minkowski spacetime, introduced by Hermann Minkowski. It is a
non-abelian Lie group with $10$ generators. Spacetime, in physical science,
single concept that recognizes the union of space and time, posited by
Albert Einstein in the theories of relativity (1905, 1916). One of the
interesting problems for Mathematicians and Physicists is. Can we do the
Fourier analysis on $P$.\ The purpose of this paper is to define the Fourier
transform in order to obtain the Plancherel formula for $G$, and then we
establish the Plancherel theorem for Spacetime(Poincare group).
\end{abstract}

\bigskip \textbf{Keywords}: Semidirect Product of Two Lie Groups, Spacetime
(Poincare Group), Fourier Transform, Plancherel Theorem

\textbf{AMS 2000 Subject Classification:} $43A30\&35D$ $05$

\section{\textbf{\ Introduction.}}

\bigskip 1. Abstract harmonic analysis is the field in which results from
Fourier analysis are extended to topological groups which are not
commutative that has connections to, theoretical physics, chemistry
analysis, algebra, geometry, and the theory of algorithms. The classical
Fourier transform is one of the most widely used mathematical tools in
engineering. However, few engineers know that extensions Fourier analysis on
noncommutative Lie groups holds great potential for solving problems in
robotics, image analysis, mechanics. Engineering applications of
noncommutative harmonic analysis brings this powerful tool to the
engineering world. The Fourier transform, known in classical analysis, and
generalized in abstract harmonic analysis. For a long time, people have
tried to construct objects in order to generalize Fourier transform and
Pontryagin,s theorem to the non abelian case. However, with the dual object
not being a group, it is not possible to define the Fourier transform and
the inverse Fourier transform between $G$ and\ $\widehat{G}.$ These
difficulties of Fourier analysis on noncommutative groups makes the
noncommutative version of the problem very challenging. It was necessary to
find a subgroup or at least a subset of locally compact groups which were
not "pathological", or "wild" as Kirillov calls them $[13$]. Unfortunately
if the group $G$ is no longer assumed to be abelian, it is not possible
anymore to consider the dual group (i.e the set of all equivalence classes
of unitary irreducible representations). Abstract harmonic analysis on
locally compact groups is generally a difficult task. Still now neither the
theory of quantum groups nor the representations theory have done to reach
this goal. Recently, these problems found a satisfactory solution with the
papers $[5,6,7,8,9].$The ways were introduced in these papers will be the
business of the expertise in the theory of abstract harmonic analysis, and
in theoretical physics. In this paper I will define the Fourier transform on
the complex semisimple Lie group $SL(2,$ $%
\mathbb{C}
),$ and then the Poincare group(Space time) $\simeq \mathbb{R}^{4}\rtimes
SL(2,$ $%
\mathbb{C}
).$

\section{Fourier Transform and Plancherel Formula on Space Time $SL(2,%
\mathbb{C}
).$}

\textbf{2. }The Lorentz discoveries on some invariance of Maxwell's
equations late in the 19 th century which were to very become the basis of
Albert Einstein's theory of special relativity. While spacetime can be
viewed as a consequence of Einstein's $1905$ theory of special relativity,
it was first explicitly proposed mathematically by one of his teachers, the
mathematician Hermann Minkowski, in a $1908$ essay building on and extending
Einstein's work. His concept of Minkowski space is the earliest treatment of
space and time as two aspects of a unified whole, the essence of special
relativity, cosmology, and gravitational fields. In the following we use the
Iwasawa decomposition of $SL(2,%
\mathbb{C}
),$ to define the Fourier transform and to demonstrate Plancherel Formula on
the complex semisimple Lie group $SL(2,%
\mathbb{C}
),.$ Therefore Let $G=SL(2,%
\mathbb{C}
)$\ be the complex Lie group, which is

\begin{equation}
SL(2,%
\mathbb{C}
)=\{\left( 
\begin{array}{cc}
a & b \\ 
c & d%
\end{array}%
\right) :\text{ }(a,b,c,d)\in 
\mathbb{C}
^{4}\text{ and }ad-bc=1\}
\end{equation}%
and let $G=KNA$ be the Iwasawa decomposition of $G$, where 
\begin{eqnarray}
K &=&K(G)=\{\left( 
\begin{array}{cc}
\alpha & \beta \\ 
\overline{\beta } & \overline{\alpha }%
\end{array}%
\right) :\text{ }(\alpha ,\beta )\in 
\mathbb{C}
^{2}\text{ \ }and\text{ \ }\left\vert \alpha \right\vert ^{2}+\left\vert
\beta \right\vert ^{2}=1\}  \notag \\
N &=&N(G)=\{\left( 
\begin{array}{cc}
1 & n \\ 
0 & 1%
\end{array}%
\right) :\text{ }n\in 
\mathbb{C}
\text{ }\}  \notag \\
A &=&A(G)=\{\left( 
\begin{array}{cc}
a & 0 \\ 
0 & a^{-1}%
\end{array}%
\right) :\text{ }a\in \mathbb{R}_{+}^{\star }\}
\end{eqnarray}

\bigskip Hence every $g\in G$ can be written as $g=kan\in G,$ where $k\in K,$
$a\in A,$ $n\in 
\mathbb{C}
.$ We denote by $L^{1}(G)$ the Banach algebra that consists of all complex
valued functions on the group $G$, which are integrable with respect to the
Haar measure of $G$ and multiplication is defined by convolution on $G$ ,
and we denote by $L^{2}(G)$ the Hilbert space of $G$. So we have for any $%
f\in L^{1}(G)$ and $\phi \in L^{1}(G)$ 
\begin{equation}
\phi \ast f(X)=\int\limits_{G}f(Y^{-1}X)\phi (Y)dY
\end{equation}

The Haar measure $dg$ on $G$ can be calculated from the Haar measures $dn;$ $%
da$ and $dk$ on $N;A$ and $K;$respectively, by the formula%
\begin{equation}
\int\limits_{G}f(g)dg=\int\limits_{A}\int\limits_{N}\int%
\limits_{K}f(ank)dadndk
\end{equation}

Keeping in mind that $a^{-2\rho }$ is the modulus of the automorphism $%
n\rightarrow $ $ana^{-1}$ of $N$ we get also the following representation of 
$dg$ 
\begin{equation}
\int\limits_{G}f(g)dg=\int\limits_{A}\int\limits_{N}\int%
\limits_{K}f(ank)dadndk=\int\limits_{N}\int\limits_{A}\int%
\limits_{K}f(nak)a^{-2\rho }dndadk
\end{equation}

Furthermore, using the relation $\int\limits_{G}f(g)dg=\int%
\limits_{G}f(g^{-1})dg,$ we receive 
\begin{equation}
\int\limits_{G}f(g)dg=\int\limits_{K}\int\limits_{A}\int%
\limits_{N}f(kan)a^{2\rho }dndadk
\end{equation}%
where $\rho =$ the dimension of $N$. Let $\underline{k}$ be the Lie algebra
of $K$ and $(X_{1},X_{2},.....,X_{m})$ a basis of $\underline{k}$ , such
that the both operators%
\begin{equation}
\Delta =\dsum\limits_{i=1}^{m}X_{i}^{2}
\end{equation}%
\begin{equation}
D_{q}=\dsum\limits_{0\leq l\leq q}\left(
-\dsum\limits_{i=1}^{m}X_{i}^{2}\right) ^{l}
\end{equation}%
are left and right invariant (bi-invariant) on $K,$ this basis exist see $%
[2, $ $p.564)$. For $l\in 
\mathbb{N}
$, let $D^{l}=(1-\Delta )^{l}$, then the family of semi-norms $\{\sigma _{l}$%
, $l\in 
\mathbb{N}
\}$ such that%
\begin{equation}
\sigma _{l}(f)=(\int_{K}\left\vert D^{l}f(y)\right\vert ^{2}dy)^{\frac{1}{2}%
},\text{ \ \ \ \ \ \ \ \ }f\in C^{\infty }(K)
\end{equation}%
define on $C^{\infty }(K)$ the same topology of the Frechet topology defined
by the semi-normas $\left\Vert X^{\alpha }f\right\Vert _{2}$ defined as%
\begin{equation}
\left\Vert X^{\alpha }f\right\Vert _{2}=(\int_{K}\left\vert X^{\alpha
}f(y)\right\vert ^{2}dy)^{\frac{1}{2}},\text{ \ \ \ \ \ \ \ \ }f\in
C^{\infty }(K)
\end{equation}%
where $\alpha =(\alpha _{1},$.....,$\alpha _{m})\in 
\mathbb{N}
^{m},$ see $[2,p.565]$

Let $\widehat{K}$ be the set of all irreducible unitary representations of $%
K.$ If $\gamma \in \widehat{K}$, we denote by $E_{\gamma }$ the space of
representation $\gamma $ and $d_{\gamma }$\ its dimension then we get\qquad

\textbf{Definition 2.1.} \textit{The Fourier transform of a function }$f\in
C^{\infty }(K)$\textit{\ is defined as} 
\begin{equation}
Tf=\dint\limits_{K}f(x)\gamma (x^{-1})dx
\end{equation}%
\textit{where }$T$\textit{\ is the Fourier transform on} $K$

\textbf{Theorem (A. Cerezo) 2.1.} \textit{Let} $f\in C^{\infty }(K),$ 
\textit{then we have the inversion of the Fourier transform} 
\begin{equation}
f(x)=\dsum\limits_{\gamma \in \widehat{K}}d\gamma tr[Tf(\gamma )\gamma (x)
\end{equation}

\begin{equation}
f(I_{K})=\dsum\limits_{\gamma \in SO(3)}d\gamma tr[Tf(\gamma )]
\end{equation}%
\textit{and the Plancherel formula} 
\begin{equation}
\left\Vert f(x)\right\Vert _{2}^{2}=\dint \left\vert f(x)\right\vert
^{2}dx=\dsum\limits_{\gamma \in \widehat{K}}d_{\gamma }\left\Vert Tf(\gamma
)\right\Vert ^{2}
\end{equation}%
\textit{for any }$f\in L^{1}(K),$ \textit{where }$I_{K}$ \textit{is the
identity element of \ }$K$

\bigskip \textbf{Definition 2.2}. \textit{For any function} $f\in \mathcal{D}%
(G),$ \textit{we can define a function} $\Upsilon (f)$\textit{on }$G\times K$
\textit{by} 
\begin{equation}
\Upsilon (f)(g,k_{1})=\Upsilon (f)(kna,k_{1})=f(gk_{1}=f(knak_{1})
\end{equation}%
\textit{for }$g=kna\in G,$ $h\in K,$ \textit{and} $k_{1}\in K$ . \textit{The
restriction of} $\ \Upsilon (f)\ast \psi (g,k_{1})$ \textit{on} $K(G)$ 
\textit{is }$\Upsilon (f)\ast \psi (g,k_{1})\downarrow _{K(G)}=f(nak_{1})\in 
\mathcal{D}(G),$ \textit{and }$\Upsilon (f)(g,k_{1})\downarrow
_{K}=f(g,I_{K})=f(kna)$ $\in \mathcal{D}(G)$

\textbf{Definition 2.3}.\textit{\ Let }$f$ \textit{and }$\psi $ \textit{be
two functions belong to} $\mathcal{D}(G),$ \textit{then we can define the
convolution of } $\Upsilon (f)\ $\textit{and} $\psi $\ \textit{on} $G$ $%
\times K$ \textit{as}

\begin{eqnarray*}
\Upsilon (f)\ast \psi (g,k_{1}) &=&\int\limits_{G}\Upsilon
(f)(gg_{2}^{-1},k_{1})\psi (g_{2})dg_{2} \\
&=&\int\limits_{K}\int\limits_{N}\int\limits_{A}\Upsilon
(f)(knaa_{2}^{-1}n_{2}^{-1}k^{-1}k_{1})\psi
(k_{2}n_{2}a_{2})dk_{2}dn_{2}da_{2}
\end{eqnarray*}%
and so we get 
\begin{eqnarray*}
\Upsilon (f)\ast \psi (g,k_{1}) &\downarrow &_{K(G)}=\Upsilon (f)\ast \psi
(I_{K}na,k_{1}) \\
&=&\int\limits_{K}\int\limits_{N}\int%
\limits_{A}f(naa_{2}^{-1}n_{2}^{-1}k^{-1}k_{1})\psi
(k_{2}n_{2}a_{2})dk_{2}dn_{2}da_{2} \\
&=&\Upsilon (f)\ast \psi (na,k_{1})
\end{eqnarray*}

\bigskip \textbf{Definition} \textbf{2.4.} \textit{If }$f\in \mathcal{D}(G)$ 
\textit{and let} $\Upsilon (f)$ \textit{be the associated function to} $f$ , 
\textit{we define the Fourier transform of \ }$\Upsilon (f)(g,k_{1})$ 
\textit{by }%
\begin{eqnarray}
&&T\mathcal{F}\Upsilon (f))(I_{K},\xi ,\lambda ,\gamma )=T\mathcal{F}%
\Upsilon (f))(I_{K},\xi ,\lambda ,\gamma )  \notag \\
&=&\int_{K}\int_{N}\int_{A}\dsum\limits_{\delta \in \widehat{K}}d_{\delta
}tr[\int_{K}\Upsilon (f)(kna,k_{1})\delta (k^{-1})dk]a^{-i\lambda }e^{-\text{
}i\langle \text{ }\xi ,\text{ }n\rangle }\text{ }\gamma
(k_{1}^{-1})dadndk_{1}  \notag \\
&=&\int_{N}\int_{A}\int_{K}\Upsilon (f)(I_{K}na,k_{1})]a^{-i\lambda }e^{-%
\text{ }i\langle \text{ }\xi ,\text{ }n\rangle }\text{ }\gamma
(k_{1}^{-1})dadndk_{1}  \notag \\
&=&\int_{N}\int_{A}\int_{K}f(I_{K}nak_{1})a^{-i\lambda }e^{-\text{ }i\langle 
\text{ }\xi ,\text{ }n\rangle }\text{ }\gamma (k_{1}^{-1})dadndk_{1}  \notag
\\
&=&\int_{N}\int_{A}\int_{K}f(nak_{1})a^{i\lambda }e^{-\text{ }i\langle \text{
}\xi ,\text{ }n\rangle }\text{ }\gamma (k_{1}^{-1})dadndk_{1}
\end{eqnarray}%
\textit{where }$\mathcal{F}$ \textit{is the Fourier transform on }$AN$ 
\textit{and }$T$ \textit{is the Fourier transform on} $K,$ \textit{and }$%
I_{K}$ \textit{is the identity element of }$K$

\textbf{\textit{\textbf{Theorem} 2.2. (Plancherel's Formula for the Group }}$%
G)$\textbf{\textit{. }}\textit{For any function\ }$f\in $\textit{\ }$%
L^{1}(G)\cap $\textit{\ }$L^{2}(G),$\textit{we get }%
\begin{eqnarray}
\int \left\vert f(g)\right\vert ^{2}dg &=&\int_{K}\int_{N}\int_{A}\left\vert
f(kna)\right\vert ^{2}dadndk  \notag \\
&=&\sum_{\delta \in \widehat{K}}d_{\gamma }\int\limits_{\mathbb{R}%
^{2}}\int\limits_{\mathbb{R}}\left\Vert T\mathcal{F}f(\lambda ,\xi ,\gamma
)\right\Vert _{2}^{2}d\lambda d\xi  \notag \\
&=&\int\limits_{\mathbb{R}^{2}}\int\limits_{\mathbb{R}}\sum_{\delta \in 
\widehat{K}}d_{\gamma }\left\Vert T\mathcal{F}f(\lambda ,\xi ,\gamma
)\right\Vert _{2}^{2}d\lambda d\xi
\end{eqnarray}%
\textit{\ }%
\begin{eqnarray}
f(I_{A}I_{N}I_{K}) &=&\int\limits_{\mathbb{R}^{2}}\int\limits_{\mathbb{R}%
}\sum_{\gamma \in \widehat{K}}d_{\gamma }tr[T\mathcal{F}f((\lambda ,\xi
,\gamma )]d\lambda d\xi \\
&=&\sum_{\gamma \in \widehat{K}}d_{\gamma }\int\limits_{\mathbb{R}%
^{2}}\int\limits_{\mathbb{R}}tr[T\mathcal{F}f(\lambda ,\xi ,\gamma
)]d\lambda d\xi =\int\limits_{\mathbb{R}^{2}}\int\limits_{\mathbb{R}%
}\sum_{\gamma \in \widehat{K}}d_{\gamma }tr[T\mathcal{F}f(\lambda ,\xi
,\gamma )]d\lambda d\xi  \notag
\end{eqnarray}%
\textit{where} $I_{A},I_{N},$ and $I_{K}$ \textit{are the identity elements
of} $A$, $N$ \textit{and }$K$ \textit{respectively, where }$\mathcal{F}$ 
\textit{is the Fourier transform on }$AN$ \textit{and }$T$ \textit{is the
Fourier transform on} $K,$ \textit{and }$I_{K}$ \textit{is the identity
element of }$K$

\bigskip \textit{Proof: }First let $\overset{\vee }{f}$ be the function
defined by 
\begin{equation}
\ \overset{\vee }{f}(kna)=\overline{f((kna)^{-1})}=\overline{%
f(a^{-1}n^{-1}k^{-1})}
\end{equation}

Then we have%
\begin{eqnarray}
&&\int \left\vert f(g)\right\vert ^{2}dg  \notag \\
&=&\mathcal{\Upsilon (}f)\ast \overset{\vee }{f}(I_{K}I_{N}I_{A},I_{K_{1}}) 
\notag \\
&=&\int\limits_{G}\Upsilon (f)(I_{K}I_{N}I_{A}(g_{2}^{-1}),I_{K_{1}})\overset%
{\vee }{f}(g_{2})dg_{2}  \notag \\
&=&\int\limits_{A}\int\limits_{N}\int\limits_{K}\Upsilon
(f)(a_{2}^{-1}n_{2}^{-1}k_{2}^{-1},I_{K})\overset{\vee }{f}%
(k_{2}n_{2}a_{2})da_{2}dn_{2}dk_{2}  \notag \\
&=&\int\limits_{A}\int\limits_{N}\int%
\limits_{K}f(a_{2}^{-1}n_{2}^{-1}k_{2}^{-1})\overline{%
f((k_{2}n_{2}a_{2})^{-1})}da_{2}dn_{2}dk_{2}  \notag \\
&=&\int\limits_{A}\int\limits_{N}\int\limits_{K}\left\vert
f(a_{2}n_{2}k_{2})\right\vert ^{2}da_{2}dn_{2}dk_{2}
\end{eqnarray}

\bigskip Secondly%
\begin{eqnarray*}
&&\Upsilon (f)\ast \overset{\vee }{f}(I_{K}I_{N}I_{A},I_{K_{1}}) \\
&=&\int\limits_{\mathbb{R}^{2}}\int\limits_{\mathbb{R}}\text{ }\mathcal{F}%
(\Upsilon (f)\ast \overset{\vee }{f})(I_{K},\lambda ,\xi ,I_{K_{1}})d\lambda
d\xi \\
&=&\int\limits_{\mathbb{R}^{2}}\int\limits_{\mathbb{R}}\sum_{\gamma \in 
\widehat{K}}d_{\gamma }\sum_{\delta \in \widehat{K}}d_{\delta }tr[T\mathcal{F%
}(\Upsilon (f)\ast \overset{\vee }{f})(\delta ,\lambda ,\xi ,\gamma
)]d\lambda d\xi \\
&=&\int\limits_{\mathbb{R}^{2}}\int\limits_{\mathbb{R}}\sum_{\gamma \in 
\widehat{K}}d_{\gamma }tr[\dint\limits_{K}\mathcal{F}(\Upsilon (f)\ast 
\overset{\vee }{f})(I_{K},\lambda ,\xi ,k_{1})\gamma
(k_{1}^{-1})dk_{1}]d\lambda d\xi \\
&=&\int\limits_{\mathbb{R}^{2}}\int\limits_{\mathbb{R}}\dint\limits_{A}\dint%
\limits_{N}\sum_{\gamma \in \widehat{K}}d_{\gamma }tr[\dint\limits_{K}%
\mathcal{F}(\Upsilon (f)\ast \overset{\vee }{f})(I_{k}na,k_{1})\gamma
(k_{1}^{-1})dk_{1}] \\
&&a^{-i\lambda }e^{-\text{ }i\langle \text{ }\xi ,\text{ }n\rangle
}dndad\lambda d\xi \\
&=&\int\limits_{\mathbb{R}^{2}}\int\limits_{\mathbb{R}}\dint\limits_{A}\dint%
\limits_{N}\dint\limits_{A}\int\limits_{N}\sum_{\gamma \in \widehat{K}%
}d_{\gamma }tr[\dint\limits_{K}\dint\limits_{K}\Upsilon
(f)(I_{k}naa_{2}^{-1}n_{2}^{-1}k_{2}^{-1},k_{1})\overset{\vee }{f}%
(k_{2}n_{2}a_{2})\gamma (k_{1}^{-1})dk_{1}dk_{2}] \\
&&a^{-i\lambda }e^{-\text{ }i\langle \text{ }\xi ,\text{ }n\rangle
}dndadn_{2}da_{2}d\lambda d\xi \\
&=&\int\limits_{\mathbb{R}^{2}}\int\limits_{\mathbb{R}}\dint\limits_{A}\dint%
\limits_{N}\dint\limits_{A}\int\limits_{N}\sum_{\gamma \in \widehat{K}%
}d_{\gamma
}tr[\dint\limits_{K}\dint\limits_{K}f(naa_{2}^{-1}n_{2}^{-1}k_{2}^{-1}k_{1})%
\overset{\vee }{f}(k_{2}n_{2}a_{2})\gamma (k_{1}^{-1})dk_{1}dk_{2}] \\
&&a^{-i\lambda }e^{-\text{ }i\langle \text{ }\xi ,\text{ }n\rangle
}dndadn_{2}da_{2}d\lambda d\xi
\end{eqnarray*}%
where%
\begin{equation}
e^{-\text{ }i\langle \text{ }\xi ,\text{ }n\rangle }=e^{-\text{ }i\langle 
\text{ }\xi ,\text{ }n\rangle }=e^{-\text{ }i\langle \text{ }(\xi _{1},\xi
_{2})\text{ (}n_{1},n_{2})\rangle }
\end{equation}

\bigskip\ Using the fact that%
\begin{equation}
\int\limits_{A}\int\limits_{N}\int\limits_{K}f(kna)dadndk=\int\limits_{N}%
\int\limits_{A}\int\limits_{K}f(kan)a^{-2}dndadk
\end{equation}%
and 
\begin{eqnarray}
&&\int\limits_{\mathbb{R}^{2}}\int\limits_{A}\int\limits_{N}\int%
\limits_{K}f(kna)e^{-\text{ }i\langle \text{ }(\xi _{1},\xi _{2})\text{ (}%
n_{1},n_{2})\rangle }dadndkd\xi _{1}d\xi _{2}  \notag \\
&=&\int\limits_{\mathbb{R}^{2}}\int\limits_{A}\int\limits_{N}\int%
\limits_{K}f(kan)e^{-\text{ }i\langle \text{ }(\xi _{1},\xi _{2}),\text{ }%
a^{-1}(n_{1},n_{2})a\text{ }\rangle }a^{-2}dadndkd\xi _{1}d\xi _{2}  \notag
\\
&=&\int\limits_{\mathbb{R}^{2}}\int\limits_{A}\int\limits_{N}\int%
\limits_{K}f(kan)e^{-\text{ }i\langle \text{ }a^{-1}(\xi _{1},\xi _{2})a,%
\text{ }(n_{1},n_{2})\rangle }a^{-2}dadndkd\xi _{1}d\xi _{2}  \notag \\
&=&\int\limits_{\mathbb{R}^{2}}\int\limits_{A}\int\limits_{N}\int%
\limits_{K}f(kan)e^{-\text{ }i\langle \text{ }(\xi _{1},\xi _{2}),\text{ }%
(n_{1},n_{2})\rangle }dadndkd\xi _{1}d\xi _{2}
\end{eqnarray}

Then we get

\begin{eqnarray*}
&&\mathcal{\Upsilon (}f)\ast \overset{\vee }{f}(I_{K}I_{N}I_{A},I_{K_{1}}) \\
&=&\int\limits_{\mathbb{R}^{2}}\int\limits_{\mathbb{R}}\text{ }\mathcal{F}%
(\Upsilon (f)\ast \overset{\vee }{f})(I_{K},\lambda ,\xi ,I_{K_{1}})d\lambda
d\xi \\
&=&\int\limits_{\mathbb{R}^{2}}\int\limits_{\mathbb{R}}\dint\limits_{A}\dint%
\limits_{N}\dint\limits_{A}\int\limits_{N}\sum_{\gamma \in \widehat{K}%
}d_{\gamma }tr[\dint\limits_{K}\dint\limits_{K}\Upsilon
(f)(I_{k}naa_{2}^{-1}n_{2}^{-1}k_{2}^{-1}k_{1})\overset{\vee }{f}%
(k_{2}n_{2}a_{2})\gamma (k_{1}^{-1})dk_{1}dk_{2}] \\
&&a^{-i\lambda }e^{-\text{ }i\langle \text{ }\xi ,\text{ }n\rangle
}dndadn_{2}da_{2}d\lambda d\xi \\
&=&\int\limits_{\mathbb{R}^{2}}\int\limits_{\mathbb{R}}\dint\limits_{A}\dint%
\limits_{N}\dint\limits_{A}\int\limits_{N}\sum_{\gamma \in \widehat{K}%
}d_{\gamma }tr[\dint\limits_{K}\dint\limits_{K}f(naa_{2}^{-1}n_{2}^{-1}k_{1})%
\overset{\vee }{f}(k_{2}n_{2}a_{2})\gamma (k_{2}^{-1})\gamma
(k_{1}^{-1})dk_{1}dk_{2}] \\
&&a^{-i\lambda }e^{-\text{ }i\langle \text{ }\xi ,\text{ }n\rangle
}dndadn_{2}da_{2}d\lambda d\xi \\
&=&\int\limits_{\mathbb{R}^{2}}\int\limits_{\mathbb{R}}\dint\limits_{A}\dint%
\limits_{N}\dint\limits_{A}\int\limits_{N}\sum_{\gamma \in \widehat{K}%
}d_{\gamma }tr[\dint\limits_{K}\dint\limits_{K}f(aa_{2}^{-1}nn_{2}^{-1}k_{1})%
\overset{\vee }{f}(k_{2}n_{2}a_{2})\gamma (k_{2}^{-1})\gamma
(k_{1}^{-1})dk_{1}dk_{2}] \\
&&a^{-i\lambda }e^{-\text{ }i\langle \text{ }\xi ,\text{ }n\rangle
}dndadn_{2}da_{2}d\lambda d\xi
\end{eqnarray*}

We continue our calcalution, we get

\begin{eqnarray*}
&&\mathcal{\Upsilon (}f)\ast \overset{\vee }{f}(I_{K}I_{N}I_{A},I_{K_{1}}) \\
&=&\int\limits_{\mathbb{R}^{2}}\int\limits_{\mathbb{R}}\dint\limits_{A}\dint%
\limits_{N}\dint\limits_{A}\int\limits_{N}\sum_{\gamma \in \widehat{K}%
}d_{\gamma }tr[\dint\limits_{K}\dint\limits_{K}f(ank_{1})\overset{\vee }{f}%
(k_{2}n_{2}a_{2})\gamma (k_{2}^{-1})\gamma (k_{1}^{-1})dk_{1}dk_{2}] \\
&&a^{-i\lambda }e^{-\text{ }i\langle \text{ }\xi ,\text{ }n\rangle
}a_{2}^{-i\lambda }e^{-\text{ }i\langle \text{ }\xi ,\text{ }n_{2}\rangle
}dndadn_{2}da_{2}d\lambda d\xi \\
&=&\int\limits_{\mathbb{R}^{2}}\int\limits_{\mathbb{R}}\dint\limits_{A}\dint%
\limits_{N}\dint\limits_{A}\int\limits_{N}\sum_{\gamma \in \widehat{K}%
}d_{\gamma }tr[\dint\limits_{K}\dint\limits_{K}f(ank_{1})\overline{%
f((k_{2}n_{2}a_{2})^{-1})}\gamma (k_{2}^{-1})\gamma (k_{1}^{-1})dk_{1}dk_{2}]
\\
&&a^{-i\lambda }e^{-\text{ }i\langle \text{ }\xi ,\text{ }n\rangle
}a_{2}^{-i\lambda }e^{-\text{ }i\langle \text{ }\xi ,\text{ }n_{2}\rangle
}dndadn_{2}da_{2}d\lambda d\xi \\
&=&\int\limits_{\mathbb{R}^{2}}\int\limits_{\mathbb{R}}\dint\limits_{A}\dint%
\limits_{N}\dint\limits_{A}\int\limits_{N}\sum_{\gamma \in \widehat{K}%
}d_{\gamma }tr[\dint\limits_{K}\dint\limits_{K}f(ank_{1})\overline{%
f((a_{2}{}^{-1}n_{2}^{-1}k_{2}^{-1})}\gamma (k_{2}^{-1})\gamma
(k_{1}^{-1})dk_{1}dk_{2}] \\
&&a^{-i\lambda }e^{-\text{ }i\langle \text{ }\xi ,\text{ }n\rangle
}a_{2}^{-i\lambda }e^{-\text{ }i\langle \text{ }\xi ,\text{ }n_{2}\rangle
}dndadn_{2}da_{2}d\lambda d\xi \\
&&\int\limits_{\mathbb{R}^{2}}\int\limits_{\mathbb{R}}\dint\limits_{A}\dint%
\limits_{N}\dint\limits_{A}\int\limits_{N}\sum_{\gamma \in \widehat{K}%
}d_{\gamma }tr[\dint\limits_{K}\dint\limits_{K}f(ank_{1})\overline{%
f((a_{2}{}n_{2}k_{2})}\gamma ^{\ast }(k_{2}^{-1})\gamma
(k_{1}^{-1})dk_{1}dk_{2}] \\
&&a^{-i\lambda }e^{-\text{ }i\langle \text{ }\xi ,\text{ }n\rangle
}a_{2}^{i\lambda }e^{\text{ }i\langle \text{ }\xi ,\text{ }n_{2}\rangle
}dndadn_{2}da_{2}d\lambda d\xi \\
&=&\int\limits_{\mathbb{R}^{2}}\int\limits_{\mathbb{R}}\sum_{\gamma \in 
\widehat{K}}d_{\gamma }T\mathcal{F}f(\lambda ,\xi ,\gamma )\overline{T%
\mathcal{F}f(\lambda ,\xi ,\gamma )}d\lambda d\xi \\
&=&\int\limits_{\mathbb{R}^{2}}\int\limits_{\mathbb{R}}\sum_{\gamma \in 
\widehat{K}}d_{\gamma }\left\Vert T\mathcal{F}(f)(\lambda ,\xi ,\gamma
)\right\Vert _{2}^{2}d\lambda d\xi
\end{eqnarray*}

\textbf{Remark 2.2. }By the same methods we can establish the Plancherel
Formula on the Lie group $SL(2,\mathbb{R})$.

\section{Fourier Transform and Plancherel Formula on Space Time $\mathbb{R}%
^{4}\rtimes SL(2,%
\mathbb{C}
).$}

\textbf{3.} To define the Fourier transform on \textbf{spacetime} (\textbf{%
Poincare group}),we need to introduce the Lorentz group $O(3,1)$. Therefore
denote the $p\times p-$identity matrix by $I_{p,p}$ and define

\begin{equation}
I_{p,q}=\left( 
\begin{array}{cc}
I_{p,p} & 0 \\ 
0 & -I_{q,q}%
\end{array}%
\right)
\end{equation}

If $n=p+q,$ the matrix $I_{p,q}$ is associated with the non degenerate
symmetric bilinear form%
\begin{equation}
\theta _{_{p,q}}((x_{1},...,x_{n}\text{ }),(y_{1},...,y_{n}\text{ }%
))=\sum_{i=1}^{p}x_{i}y_{i}-\sum_{j=1}^{q}x_{i}y_{i}
\end{equation}%
with associated quadratic form$\ \ \ \ \ \ \ \ \ \ \ \ \ \ \ \ \ \ \ \ \ \ \
\ \ \ \ \ \ \ \ \ \ \ \ \ \ \ \ \ \ \ \ \ \ \ \ \ \ \ \ \ \ \ \ \ \ \ \ \ \
\ \ \ \ \ \ \ \ \ \ \ \ \ \ \ \ \ \ \ \ \ \ \ \ \ \ \ \ \ \ \ \ \ \ \ \ \ \
\ \ \ $%
\begin{equation}
\theta
_{p,q}((x_{1},...,x_{n}))=\sum_{i=1}^{p}x_{i}^{2}-\sum_{j=1}^{q}x_{j}^{2}
\end{equation}

We denote by $O(p,q)$ the group consisting of all matrices of the form%
\begin{equation}
O(p,\text{ }q\mathbb{)}=\{A\in GL(n,\mathbb{R)},\text{ }A^{t}I_{p,q}A=I_{p,q}%
\text{ }\}
\end{equation}%
which is the group of isometries $\theta _{p,q}$,where $A^{t}$ is the
transpose matrix of $A.$ the group$O(p,q)$ has a sub-group consisting of all
matrices with determinant $+1$ and will be denoted $SO(p,$ $q\mathbb{)}$
that is 
\begin{equation}
SO(p,\text{ }q\mathbb{)}=\{A\in O(p,q\mathbb{)},\text{ }\det \text{ }A=1\}
\end{equation}

Since our interest is the Lorentz group, then we will restrict our attention
to the case for $p=3$ and $q=1$consider, $i.e$\ the group $O(3,$ $1\mathbb{)}
$, which is called sometimes the full Lorentz group preserve the quadratic
form 
\begin{equation}
\theta _{3,1}((x_{1},\text{ }x_{2},\text{ }x_{3},\text{ }x_{4}))=\dsum%
\limits_{i=1}^{3}x_{i}^{2}-x_{4}^{2}
\end{equation}

In physics, $x_{1}$ is interpreted as time and written $t$ and $%
x_{2},x_{3},x_{4}$ as coordinates in $\mathbb{R}^{3}$ and written $x,y,z$.
Thus, the Lorentz metric is usually written a $t^{2}-$ $x^{2}-y^{2}-z^{2}$
although it also appears as $x^{2}+y^{2}+z^{2}-t^{2}.$ The space $\mathbb{R}%
^{4}$ with the Lorentz metric is called \textit{Minkowski space. }It plays
an important role in Einstein,s theory of special relativity. The group $%
O(3,1)$ as a manifold has four connected components,we denote by $%
SO_{+}(3,1) $ the connected component of the identity. The Poincare group
(spacetime) is given by

\begin{eqnarray}
P &=&\mathbb{R}^{4}\rtimes SO_{+}(1;3)\text{ }or\text{ }with\text{ }%
double-cover  \notag \\
\text{ }\widetilde{P} &=&\mathbb{R}^{4}\rtimes SL(2;%
\mathbb{C}
)=\mathbb{R}^{4}\rtimes G
\end{eqnarray}

The action of $SL(2;%
\mathbb{C}
)$ on $\mathbb{R}^{4}$ is the action of the Lorentz group on Minkowski
space. We can define the action of $SL(2;%
\mathbb{C}
)$ on Minkowski spacetime by writing a point of spacetime as a two-by-two
Hermitian matrix in the form%
\begin{equation}
M=\left( 
\begin{array}{cc}
t+z & x-iy \\ 
x+y & t-z%
\end{array}%
\right) :\text{ }(x,y,t,z)\in \mathbb{R}^{4}\text{\ }
\end{equation}

This presentation has the pleasant feature that%
\begin{equation}
\det M=t^{2}-x^{2}-y^{2}-z^{2}\text{\ }
\end{equation}

Therefore, we have identified the space of Hermitian matrices (which is four
dimensional, as a real vector space) with Minkowski spacetime in such a way
that the determinant of a Hermitian matrix $M$ is the squared length of the
corresponding vector in Minkowski spacetime. $SL(2;%
\mathbb{C}
)$ acts on the space of Hermitian matrices via the group homomorphism $\rho
:SL(2;%
\mathbb{C}
)\rightarrow Aut(\mathbb{R}^{4})$ defined by: 
\begin{equation}
\rho (g)(M)=gMg^{\star }=M,\text{ \ }g\in G
\end{equation}%
where $Aut(\mathbb{R}^{4})$ is the group of all automorphisms of $\mathbb{R}%
^{4},$ $g^{\star }$ is the Hermitian transpose of $g$, and this action
preserves the determinant. Therefore, $SL(2;%
\mathbb{C}
)$ acts on Minkowski spacetime by (linear) isometries, and so is isomorphic
to a subset of the Lorentz group by the definition of the Lorentz group.
Multiplication in $P$ is then given as 
\begin{equation}
(v,g)(v^{\prime },g^{\prime })=(v+\text{\ }\rho (g)(v^{\prime }),\text{ }%
gg^{\prime })=(v+\text{\ }gv^{\prime },\text{ }gg^{\prime })\text{ \ }
\end{equation}%
for any $(v,v^{\prime })\in \mathbb{R}^{4}\times \mathbb{R}^{4}$ and $(g,$ $%
g^{\prime })\in G\times G,$ where $gv^{\prime }=\rho (g)(v^{\prime }).$ To
define the Fourier transform on spacetime, we introduce the following new
group

\textbf{Definition 3.1}. Let $Q=\mathbb{R}^{4}$ $\times G\times G$ be the
group with law:%
\begin{eqnarray}
X\cdot Y &=&(v,h,g)(v^{\prime },h^{\prime },g^{\prime })  \notag \\
&=&(v+\text{\ }gv^{\prime },hh^{\prime },\text{ }gg^{\prime })
\end{eqnarray}%
for all $X=(v,h,g)$ $\in G$ and $Y=(v^{\prime },h^{\prime },g^{\prime })\in
G $

\textbf{Definition 3.2}. \textit{For any function} $f\in \mathcal{D}(P),$ 
\textit{we can define a function} $\widetilde{f}$ \textit{on }$Q$ \textit{by}

\begin{equation}
\widetilde{f}(v,g,h)=f(gv,gh)
\end{equation}

\textbf{Remark 3.1. }The function $\widetilde{f}$ \ is invariant in the
following sense%
\begin{equation}
\widetilde{f}(q^{-1}v,g,q^{-1}h)=\widetilde{f}(v,gq^{-1},h)\text{\ }
\end{equation}

\textbf{Theorem 3.1. }\textit{For any function }$\psi \in \mathcal{D}(P)$ 
\textit{and }$\widetilde{f}\in \mathcal{D}(Q)$ \textit{invariant in sense }$%
(36)$\textit{, we get}%
\begin{equation}
\psi \ast \widetilde{f}(v,h,g)=\psi \ast _{c}\widetilde{f}(v,h,g)
\end{equation}%
\textit{\ }and\textit{where} $\ast $ \textit{signifies the convolution
product on} $P$ \textit{with respect the variable} $(v,h),$ \textit{and }$%
\ast _{c}$\textit{signifies the convolution product on} $A$ \textit{with
respect the variable} $(v,g)$

\textit{Proof : }In fact we have%
\begin{eqnarray}
&&\psi \ast \widetilde{f}(v,h,g)  \notag \\
&=&\int\limits_{\mathbb{R}^{4}}\int_{G}\widetilde{f}((v^{\prime },g^{\prime
})^{-1}(v,h,g))\psi (v^{\prime },g^{\prime })dv^{\prime }dg^{\prime }  \notag
\\
&=&\int\limits_{\mathbb{R}^{4}}\int_{G}\widetilde{f}[(g^{\prime
}{}^{-1}(-v^{\prime }),g^{\prime }{}^{-1})(v,h,g)]\psi (v^{\prime
},g^{\prime })dv^{\prime }dg^{\prime }  \notag \\
&=&\int\limits_{\mathbb{R}^{4}}\int_{G}\widetilde{f}[(g^{\prime
}{}^{-1}(-v^{\prime }),g^{\prime }{}^{-1})(v,h,g)]\psi (v^{\prime
},g^{\prime })dv^{\prime }dg^{\prime }  \notag \\
&=&\int\limits_{\mathbb{R}^{4}}\int_{G}\widetilde{f}[(g^{\prime
}{}^{-1}(v-v^{\prime }),h,g^{\prime }{}^{-1}g)]\psi (v^{\prime },g^{\prime
})dv^{\prime }dg^{\prime }  \notag \\
&=&\int\limits_{\mathbb{R}^{4}}\int_{G}\widetilde{f}[v-v^{\prime
},hg^{\prime }{}^{-1},g]\psi (v^{\prime },g^{\prime })dv^{\prime }dg^{\prime
}  \notag \\
&=&\psi \ast _{c}\widetilde{f}(v,h,g)
\end{eqnarray}%
\textbf{\ }

\textbf{Definition 3}.\textbf{3}. \textit{Let }$\Upsilon F$ \textit{be the
function on }$P\times K$ \textit{defined by}%
\begin{equation}
\Upsilon F(v,(g,k_{1}))=F(v,gk_{1})
\end{equation}%
\textit{and let} $h(F)$ \textit{be the function on }$P$ \textit{defined by}%
\begin{equation}
h(F)(v,g)=F(gv,g)
\end{equation}

\textbf{Definition 3.4. \ }\textit{Let }$\psi \in \mathcal{D}(P)$ \textit{%
and }$F\in \mathcal{D}(P),$\textit{then we can define a convolution product
on the Poincare group }$P$ \textit{as}%
\begin{eqnarray*}
\psi \ast _{c}\Upsilon F(v,(g,k_{1})) &=&\int\limits_{\mathbb{R}%
^{4}}\int_{G}\Upsilon F(v-v^{\prime },(gg^{\prime }{}^{-1},k_{1}))\psi
(v^{\prime },g^{\prime })dv^{\prime }dg^{\prime } \\
&=&\int\limits_{\mathbb{R}^{4}}\int_{K}\int_{N}\int_{A}F(v-v^{\prime
},kna(k^{\prime }n^{\prime }a^{\prime })^{-1}k_{1}))\psi (v^{\prime
},k^{\prime }n^{\prime }a^{\prime })dv^{\prime }dk^{\prime }dn^{\prime
}da^{\prime }
\end{eqnarray*}%
\textit{where }$g=kna$ \textit{and} $g^{\prime }=k^{\prime }n^{\prime
}a^{\prime }$

\textbf{Corollary 3.1. }\textit{For any function }$F$ \textit{belongs to} $%
\mathcal{D}(P)$ , \textit{we obtain}%
\begin{eqnarray}
\psi \ast _{c}\Upsilon h(F)(v,(g,k_{1})) &=&\int\limits_{\mathbb{R}%
^{4}}\int_{G}\Upsilon h(F)(v-v^{\prime },(gg^{\prime }{}^{-1},k_{1})\psi
(v^{\prime },g^{\prime })dv^{\prime }dg^{\prime }  \notag \\
&=&\int\limits_{\mathbb{R}^{4}}\int_{G}\Upsilon h(F)(v-v^{\prime
},(gg^{\prime }{}^{-1},k_{1})\psi (v^{\prime },g^{\prime })dv^{\prime
}dg^{\prime }  \notag \\
&=&\int\limits_{\mathbb{R}^{4}}\int_{G}h(F)(v-v^{\prime },gg^{\prime
}{}^{-1}k_{1})\psi (v^{\prime },g^{\prime })dv^{\prime }dg^{\prime }  \notag
\\
&=&\int\limits_{\mathbb{R}^{4}}\int_{G}F(gg^{\prime
}{}^{-1}k_{1}(v-v^{\prime }),gg^{\prime }{}^{-1}k_{1})\psi (v^{\prime
},g^{\prime })dv^{\prime }dg^{\prime }
\end{eqnarray}

\textbf{Corollary 3.2. }\textit{For any function }$F$ \textit{belongs to} $%
\mathcal{D}(P)$ , \textit{we obtain}\textbf{\ } 
\begin{equation}
F\ast \Upsilon h(\overset{\vee }{F})(0,(I_{G},I_{K_{1}}))=\int\limits_{G}%
\dint\limits_{\mathbb{R}^{4}}\left\Vert f(v,g)\right\Vert
^{2}dgdv=\left\Vert f\right\Vert _{2}^{2}
\end{equation}

\textit{Proof: }If\ $F\in \mathcal{D}(P),$then we get

\begin{eqnarray*}
&&F\ast \Upsilon h(\overset{\vee }{F})(0,(I_{G},I_{K_{1}})) \\
&=&\int\limits_{G}\dint\limits_{\mathbb{R}^{4}}\mathit{\ }\Upsilon \hbar (%
\overset{\vee }{F})[(0-v),(I_{G}g^{-1},I_{K_{1}})]F(v,g)dgdv \\
&=&\int\limits_{G}\dint\limits_{\mathbb{R}^{4}}\mathit{\ }\hbar (\overset{%
\vee }{F})[(0-v),I_{G}g^{-1}I_{K_{1}}]F(v,g)dgdv \\
&=&\int\limits_{G}\dint\limits_{\mathbb{R}^{4}}\mathit{\ }\hbar (\overset{%
\vee }{F})[(-v),g^{-1}]F(v,g)dgdv \\
&=&\int\limits_{G}\dint\limits_{\mathbb{R}^{4}}\mathit{\ }\overset{\vee }{F}%
[g^{-1}(-v),g^{-1}]F(v,g)dgdv \\
&=&\int\limits_{G}\dint\limits_{\mathbb{R}^{4}}\overline{%
F[g^{-1}(-v),g^{-1}]^{-1}}F(v,g)dgdv=\int\limits_{G}\dint\limits_{\mathbb{R}%
^{4}}\overline{F[v,g]}F(v,g)dgdv \\
&=&\int\limits_{G}\dint\limits_{\mathbb{R}^{4}}\left\Vert f(v,g)\right\Vert
^{2}dgdv=\left\Vert f\right\Vert _{2}^{2}
\end{eqnarray*}

\textbf{Definition 3.4.}\textit{\ Let} $f\in \mathcal{D}(P),$ \textit{we
define its Fourier transform by}%
\begin{equation}
\mathcal{F}_{\mathbb{R}^{4}}T\mathcal{F}f(\eta ,\gamma ,\xi ,\lambda
)=\int\limits_{\mathbb{R}^{4}}\int_{A}\int_{N}\int_{K}f(v,kna)e^{-\text{ }%
i\langle \text{ }\eta ,\text{ }v\rangle }\text{ }\gamma (k^{-1})a^{-i\lambda
}e^{-\text{ }i\langle \text{ }\xi ,\text{ }n\rangle }dkdadnd\lambda d\xi dv 
\notag
\end{equation}%
\textit{where} $\mathcal{F}_{\mathbb{R}^{4}}$ \textit{is the Fourier
transform on} $\mathbb{R}^{4},$ $kna=g,$ $\eta =(\eta _{1},\eta _{2},\eta
_{3},\eta _{4})\in \mathbb{R}^{4},$ $v=(v_{1},v_{2},v_{3},v_{4})\in \mathbb{R%
}^{4},$ \textit{and} $dv=dv_{1}dv_{2}dv_{3}dv_{4}$ \textit{is the Lebesgue
measure on} $\mathbb{R}^{4}$

\begin{eqnarray}
\langle \eta ,v\rangle &=&\langle (\eta _{1},\eta _{2},\eta _{3},\eta
_{4}),(v_{1},v_{2},v_{3},v_{4})\rangle  \notag \\
&=&\eta _{1}v_{1}+\eta _{2}v_{2}+v_{3}\eta _{3}+\eta _{4}v_{4}
\end{eqnarray}

\bigskip To obtain the Plancherel formula for the space time, we refer to $%
[8,9]$

\textbf{\textit{Plancherel's Theorem for spacetime \textbf{3.2}. }}\textit{%
For any function }$f\in $\textit{\ }$L^{1}(P)\cap $\textit{\ }$L^{2}(P),$%
\textit{we get}%
\begin{equation}
\int_{p}\left\vert f(v,g)\right\vert ^{2}dvdg=\int\limits_{\mathbb{R}%
^{2}}\dint\limits_{\mathbb{R}}\dint\limits_{\mathbb{R}^{4}}\sum_{\gamma \in 
\widehat{K}}d_{\gamma }tr\left\Vert \mathcal{F}_{\mathbb{R}^{4}}T\mathcal{F}%
F(\eta ,\gamma ,\xi ,\lambda )\right\Vert ^{2}d\eta d\lambda d\xi
\end{equation}

\textit{Proof: }Let\ $\Upsilon \hbar (\overset{\vee }{F})$ be the function
defined as\textit{\ }%
\begin{eqnarray}
&&\mathit{\ }\Upsilon \hbar (\overset{\vee }{F})\text{\ }(v;(g,k_{1}))=%
\mathit{\ }\hbar (\overset{\vee }{F})\text{\ }(v;gk_{1})  \notag \\
&=&\overset{\vee }{F}(gk_{1}v;gk_{1})=\mathit{\ }\overline{%
F(gk_{1}v;gk_{1})^{-1})}
\end{eqnarray}%
then, we have

\begin{eqnarray*}
&&F\ast \Upsilon \mathit{\ }\hbar (\overset{\vee }{F})\text{\ }%
(0,(I_{G},I_{K_{1}})) \\
&=&F\ast \Upsilon \mathit{\ }\hbar (\overset{\vee }{F})\text{\ }%
(0,(I_{K}I_{N}I_{A},I_{K_{1}})) \\
&=&\int\limits_{\mathbb{R}^{2}}\dint\limits_{\mathbb{R}}\dint\limits_{%
\mathbb{R}^{4}}\mathcal{F}_{\mathbb{R}^{4}}\mathcal{F[}F\ast \Upsilon 
\mathit{\ }\hbar (\overset{\vee }{F}\text{\ }](\eta ,(I_{K},\xi ,\lambda
,I_{K_{1}}))d\eta d\lambda d\xi \\
&=&\int\limits_{\mathbb{R}^{2}}\dint\limits_{\mathbb{R}}\dint\limits_{%
\mathbb{R}^{4}}\sum_{\gamma \in \widehat{K}}d_{\gamma }tr[\dint\limits_{K}%
\mathcal{F}_{\mathbb{R}^{4}}T\mathcal{F(}F\ast \Upsilon \mathit{\ }\hbar (%
\overset{\vee }{F}))((\eta ,(I_{K},\xi ,\lambda ,k_{1}))\gamma
(k_{1}^{-1})dk_{1})]d\eta d\lambda d\xi \\
&=&\int\limits_{\mathbb{R}^{2}}\dint\limits_{\mathbb{R}}\dint\limits_{%
\mathbb{R}^{4}}\int\limits_{\mathbb{R}^{2}}\dint\limits_{\mathbb{R}%
}\dint\limits_{\mathbb{R}^{4}}\mathit{\ }\sum_{\gamma \in \widehat{K}%
}d_{\gamma }tr[\dint\limits_{K}\mathcal{(}F\ast \Upsilon \mathit{\ }\hbar (%
\overset{\vee }{F}))((v,(I_{K}na,k_{1}))\gamma (k_{1}^{-1})dk_{1})] \\
&&e^{-\text{ }i\langle \text{ }\eta ,\text{ }v\rangle }a^{-i\lambda }e^{-%
\text{ }i\langle \text{ }\xi ,\text{ }n\rangle }dadndvd\eta d\lambda d\xi \\
&=&\dint\limits_{G}\int\limits_{\mathbb{R}^{2}}\dint\limits_{\mathbb{R}%
}\dint\limits_{\mathbb{R}^{4}}\int\limits_{\mathbb{R}^{2}}\dint\limits_{%
\mathbb{R}}\dint\limits_{\mathbb{R}^{4}}\sum_{\gamma \in \widehat{K}%
}d_{\gamma }tr[\dint\limits_{K}\Upsilon \mathit{\ }\hbar (\overset{\vee }{F}%
))((v-w),(I_{K}nag_{2}^{-1},k_{1}))\gamma (k_{1}^{-1})dk_{1})] \\
&&F(w,g_{2})e^{-\text{ }i\langle \text{ }\eta ,\text{ }v\rangle
}a^{-i\lambda }e^{-\text{ }i\langle \text{ }\xi ,\text{ }n\rangle
}dg_{2}dadndvd\eta d\lambda d\xi \\
&=&\int\limits_{\mathbb{R}^{2}}\dint\limits_{\mathbb{R}}\dint\limits_{%
\mathbb{R}}\dint\limits_{\mathbb{R}^{4}}\int\limits_{\mathbb{R}%
^{2}}\int\limits_{\mathbb{R}^{2}}\dint\limits_{\mathbb{R}}\dint\limits_{%
\mathbb{R}^{4}}\dint\limits_{\mathbb{R}^{4}}\dint\limits_{K}\sum_{\gamma \in 
\widehat{K}}d_{\gamma }tr[\mathit{\ }\dint\limits_{K}\hbar (\overset{\vee }{F%
}))((v-w),(naa_{2}^{-1}n_{2}^{-1}k_{2}^{-1}k_{1}))\gamma
(k_{1}^{-1})dk_{1})dk_{2}] \\
&&F(w,k_{2}n_{2}a_{2})e^{-\text{ }i\langle \text{ }\eta ,\text{ }v\rangle
}a^{-i\lambda }e^{-\text{ }i\langle \text{ }\xi ,\text{ }n\rangle
}da_{2}dn_{2}dadndwdvd\eta d\lambda d\xi \\
&=&\int\limits_{\mathbb{R}^{2}}\dint\limits_{\mathbb{R}}\dint\limits_{%
\mathbb{R}}\dint\limits_{\mathbb{R}^{4}}\int\limits_{\mathbb{R}%
^{2}}\int\limits_{\mathbb{R}^{2}}\dint\limits_{\mathbb{R}}\dint\limits_{%
\mathbb{R}^{4}}\dint\limits_{\mathbb{R}^{4}}\sum_{\gamma \in \widehat{K}%
}d_{\gamma }tr[\mathit{\ }\dint\limits_{K}\dint\limits_{K}\hbar (\overset{%
\vee }{F}))((v,(ank_{1}))\gamma
(k_{1}^{-1})dk_{1})F(w,k_{2}n_{2}a_{2})\gamma (k_{2}^{-1})dk_{2}] \\
&&e^{-\text{ }i\langle \text{ }\eta ,\text{ }v+w\rangle }a^{-i\lambda
}a_{2}^{-i\lambda }e^{-\text{ }i\langle \text{ }\xi ,\text{ }n+n_{2}\rangle
}da_{2}dn_{2}dadndwdvd\eta d\lambda d\xi \\
&=&\int\limits_{\mathbb{R}^{2}}\dint\limits_{\mathbb{R}}\dint\limits_{%
\mathbb{R}}\dint\limits_{\mathbb{R}^{4}}\int\limits_{\mathbb{R}%
^{2}}\int\limits_{\mathbb{R}^{2}}\dint\limits_{\mathbb{R}}\dint\limits_{%
\mathbb{R}^{4}}\dint\limits_{\mathbb{R}^{4}}\sum_{\gamma \in \widehat{K_{1}}%
}d_{\gamma }tr[\mathit{\ }\hbar (\overset{\vee }{F}%
)(v,(ank_{1}))F(w,k_{2}n_{2}a_{2})\gamma (k_{1}^{-1})\gamma
(k_{2}^{-1})dk_{1}dk_{2}] \\
&&e^{-\text{ }i\langle \text{ }\eta ,\text{ }v+w\rangle }a^{-i\lambda
}a_{2}^{-i\lambda }e^{-\text{ }i\langle \text{ }\xi ,\text{ }n+n_{2}\rangle
}da_{2}dn_{2}dadndwdvd\eta d\lambda d\xi
\end{eqnarray*}

Continuing calculation we get%
\begin{eqnarray*}
&&F\ast \Upsilon \mathit{\ }\hbar (\overset{\vee }{F})\text{\ }%
(0,(I_{K}I_{N}I_{A},I_{K_{1}})) \\
&=&\int\limits_{\mathbb{R}^{2}}\dint\limits_{\mathbb{R}}\dint\limits_{%
\mathbb{R}}\dint\limits_{\mathbb{R}^{4}}\int\limits_{\mathbb{R}%
^{2}}\int\limits_{\mathbb{R}^{2}}\dint\limits_{\mathbb{R}}\dint\limits_{%
\mathbb{R}^{4}}\dint\limits_{\mathbb{R}^{4}}\sum_{\gamma \in \widehat{K}%
}d_{\gamma }tr[\dint\limits_{K}\dint\limits_{K}(\overset{\vee }{F}%
)(ank_{1}v,ank_{1})F(w,k_{2}n_{2}a_{2})\gamma (k_{1}^{-1})\gamma
(k_{2}^{-1})dk_{1}dk_{2}] \\
&&e^{-\text{ }i\langle \text{ }\eta ,\text{ }v\rangle }e^{-\text{ }i\langle 
\text{ }\eta ,\text{ }w\rangle }a^{-i\lambda }a_{2}^{-i\lambda }e^{-\text{ }%
i\langle \text{ }\xi ,\text{ }n\rangle }e^{-\text{ }i\langle \text{ }\xi ,%
\text{ }n_{2}\rangle }da_{2}dn_{2}dadndwdvd\eta d\lambda d\xi \\
&=&\int\limits_{\mathbb{R}^{2}}\dint\limits_{\mathbb{R}}\dint\limits_{%
\mathbb{R}}\dint\limits_{\mathbb{R}^{4}}\int\limits_{\mathbb{R}%
^{2}}\int\limits_{\mathbb{R}^{2}}\dint\limits_{\mathbb{R}}\dint\limits_{%
\mathbb{R}^{4}}\dint\limits_{\mathbb{R}^{4}}\sum_{\gamma \in \widehat{K}%
}d_{\gamma }tr[\dint\limits_{K}\dint\limits_{K}\overline{%
F(ank_{1}v,ank_{1})^{-1}}F(w,k_{2}n_{2}a_{2})\gamma (k_{1}^{-1})\gamma
(k_{2}^{-1})dk_{1}dk_{2}] \\
&&e^{-\text{ }i\langle \text{ }\eta ,\text{ }v\rangle }e^{-\text{ }i\langle 
\text{ }\eta ,\text{ }w\rangle }a^{-i\lambda }a_{2}^{-i\lambda }e^{-\text{ }%
i\langle \text{ }\xi ,\text{ }n\rangle }e^{-\text{ }i\langle \text{ }\xi ,%
\text{ }n_{2}\rangle }da_{2}dn_{2}dadndwdvd\eta d\lambda d\xi \\
&=&\int\limits_{\mathbb{R}^{2}}\dint\limits_{\mathbb{R}}\dint\limits_{%
\mathbb{R}}\dint\limits_{\mathbb{R}^{4}}\int\limits_{\mathbb{R}%
^{2}}\int\limits_{\mathbb{R}^{2}}\dint\limits_{\mathbb{R}}\dint\limits_{%
\mathbb{R}^{4}}\dint\limits_{\mathbb{R}^{4}}\sum_{\gamma \in \widehat{K}%
}d_{\gamma }tr[\dint\limits_{K}\dint\limits_{K}\overline{%
F(-v,k_{1}^{-1}n^{-1}a^{-1})}F(w,k_{2}n_{2}a_{2})\gamma (k_{1}^{-1})\gamma
(k_{2}^{-1})dk_{1}dk_{2}] \\
&&e^{-\text{ }i\langle \text{ }\eta ,\text{ }v\rangle }e^{-\text{ }i\langle 
\text{ }\eta ,\text{ }w\rangle }a^{-i\lambda }a_{2}^{-i\lambda }e^{-\text{ }%
i\langle \text{ }\xi ,\text{ }n\rangle }e^{-\text{ }i\langle \text{ }\xi ,%
\text{ }n_{2}\rangle }da_{2}dn_{2}dadndwdvd\eta d\lambda d\xi \\
&=&\int\limits_{\mathbb{R}^{2}}\dint\limits_{\mathbb{R}}\dint\limits_{%
\mathbb{R}}\dint\limits_{\mathbb{R}^{4}}\int\limits_{\mathbb{R}%
^{2}}\int\limits_{\mathbb{R}^{2}}\dint\limits_{\mathbb{R}}\dint\limits_{%
\mathbb{R}^{4}}\dint\limits_{\mathbb{R}^{4}}\sum_{\gamma \in \widehat{K}%
}d_{\gamma }tr[\dint\limits_{K}\dint\limits_{K}\overline{F(v,k_{1}na)}%
F(w,k_{2}n_{2}a_{2})\gamma ^{\ast }(k_{1}^{-1})\gamma
(k_{2}^{-1})dk_{1}dk_{2}] \\
&&e^{\text{ }i\langle \text{ }\eta ,\text{ }v\rangle }e^{-\text{ }i\langle 
\text{ }\eta ,\text{ }w\rangle }a^{i\lambda }a_{2}^{-i\lambda }e^{\text{ }%
i\langle \text{ }\xi ,\text{ }n\rangle }e^{-\text{ }i\langle \text{ }\xi ,%
\text{ }n_{2}\rangle }da_{2}dn_{2}dadndwdvd\eta d\lambda d\xi \\
&=&\int\limits_{\mathbb{R}^{2}}\dint\limits_{\mathbb{R}}\dint\limits_{%
\mathbb{R}^{4}}\sum_{\gamma \in \widehat{K}}d_{\gamma }tr\mathit{\ }%
\overline{\mathcal{F}_{\mathbb{R}^{4}}T\mathcal{F[}F\text{\ }(\eta ,\gamma
^{\ast },\xi ,\lambda )}\mathcal{F}_{\mathbb{R}^{4}}T\mathcal{F[}F\text{\ }%
(\eta ,\gamma ,\xi ,\lambda )d\eta d\lambda d\xi \\
&=&\int\limits_{\mathbb{R}^{2}}\dint\limits_{\mathbb{R}}\dint\limits_{%
\mathbb{R}^{4}}\sum_{\gamma \in \widehat{K}}d_{\gamma }tr\mathit{\ }%
\left\Vert \mathcal{F}_{\mathbb{R}^{4}}T\mathcal{F[}F\text{\ }(\eta ,\gamma
,\xi ,\lambda )\right\Vert ^{2}d\eta d\lambda d\xi
\end{eqnarray*}%
where $(0,0,0,0)$ is the identity element of the vector group $\mathbb{R}%
^{4} $ and $I_{K\text{ }}I_{N}I_{A}$ is the identity element of the complex
semisimple Lie group $SL(2;%
\mathbb{C}
)=KNA.$

\section{\protect\bigskip Conclusion.}

Combining these results with the results were obtained in $[8,9]$ help us to
define the Fourier transform for the Lie groups $SL(n$, $\mathbb{R}),$ $SL(n$%
, $\mathbb{%
\mathbb{C}
}),$ $GL_{+}(n$, $\mathbb{R}),$ $GL_{-}(n$, $\mathbb{R}),$ and for the
affine group $\mathbb{R}^{n}\rtimes $ $GL(n$, $\mathbb{R}).$

\end{document}